\documentclass[sn-mathphys,Numbered]{sn-jnl}

\usepackage[version-1-compatibility]{siunitx}

\usepackage{physics}
\usepackage{graphicx}%
\usepackage{multirow}%
\usepackage{amsmath,amssymb,amsfonts}%
\usepackage{amsthm}%
\usepackage{mathrsfs}%
\usepackage[title]{appendix}%
\usepackage{xcolor}%
\usepackage{textcomp}%
\usepackage{manyfoot}%
\usepackage{booktabs}%
\usepackage{algorithm}%
\usepackage{algorithmicx}%
\usepackage{algpseudocode}%
\usepackage{listings}%

\theoremstyle{thmstyleone}%
\theoremstyle{thmstyletwo}%

\theoremstyle{thmstylethree}%

\usepackage{csquotes}
\usepackage{tensor}
\begin{document}

\title[Article Title]{On a theory of isotopic spin\footnote{[English translation of: P. Gulmanelli, \textit{Su una teoria dello spin isotopico}. Pubblicazioni della Sezione di Milano dell'Istituto Nazionale di Fisica Nucleare (Casa Editrice Pleion, Milano, 1957). Translated by A. E. S. Hartmann, who is particularly grateful to Giovanna Colombo (Biblioteca di Scienza dell'Universit{\`a} degli Studi della Insubria, Como) for finding a copy of the original notes, provided by the Biblioteca delle Scienze dell'Universit{\`a} degli Studi di Pavia. (23 October 2023)]}}


\author[1,2]{\fnm{Paolo} \sur{Gulmanelli}\footnote{Forl{\`i}, 1928 -- Pavia, 2017.}\; }



\affil[1]{\orgdiv{Istituto di Scienze Fisiche}, \orgname{Universit{\`a} degli Studi di Milano},  \country{Italy}}

\affil[2]{ \orgname{INFN, Sez. Milano}}



\abstract{The arguments presented here were the subject of a series of seminars at the Institute of Theoretical Physics of the University of Milan. It concerns a study carried on, in late 1953, at the ETH Zurich and related to a theory developed by Prof. W. Pauli.\\
Since the author has not published the present theory in any journal, the writer considered opportune to give also an extended exposition (section II, until paragraph g).}

\maketitle

\section{Isotopic spin and quantization of mass} 

It is known that if one wants an irreducible wave equation, describing the proton and the neutron at once, that is, if one wants to introduce, in an irreducible way, a vector matrix $\vec{\tau}$ as a further specification of the properties of particles with spin $1/2$, one needs to enlarge the space of the ordinary chronotope. 

Pais (\textit{Physica} \href{https://doi.org/10.1016/S0031-8914(53)80098-4}{\textbf{19}, 869, 1953}) suggested to treat the chronotopic point as a manifold ($\omega-$space) invariant under all transformations of the 3-dimensional real orthogonal group (that is, as the surface of a 3-dimensional sphere, for instance).

The results are: 1) the independence of the charge of nuclear forces follows directly from the invariance of the interaction term with respect to the rotations of the $\omega-$space; 2) the \textit{iterate} wave equations, which are satisfied by a 8-component spinor, provide a mass spectrum whose levels are dependent on the quantum numbers related to the new degrees of freedom of the particle. The lowest level corresponds to the nucleons. 

Pauli remarked that, in the framework of Pais, 1) the (pseudoscalar) tensorial character of the meson field do not arise from the theory for any intrinsic reason, but it is postulated; 2) the charge conservation (those operators that corresponds to the observable quantities are obviously expressed in terms of integrals over the variables of the $\omega-$space, and hence are explicitly independent) is related to a transformation of constant phase. It would be interesting, instead, to deduce it from a more general aspect, whose analogue would be the gauge transformations of the electromagnetic potentials. The criteria, which was Pauli's inspiration for the transformations in his theory, is the one of unitary theories, to attribute to the space a structure from which results possible to deduce the existence of forces susceptible to physical interpretation. In our case, the forces are nuclear and the potentials under search are the ones of the meson field.

\section{The transformations group of the $\omega-$space and the metric field}

At each point of the spacetime of general relativity, it is postulated the existence of a n-dimensional $\omega-$space.

The connection between the two spaces ([which are] totally independent in Pais theory) is introduced by postulating that the rotations in $\omega$ are point dependent.

The group of homogeneous transformations to be taken is the following:
\begin{equation}\label{Eq1}
\begin{split}
x'^{i} &= x^i \quad (i=1,2,3,4) \\
\Omega'^{A} & = \tensor{a}{^A_B}(x)\, \Omega^B \quad (A,B = 1,2,...,n) \\
\Omega^B &= \tensor{\overline{a}}{_C^B} (x')\, \Omega'^C,
\end{split}
\end{equation}
with $\tensor{a}{^A_B}\,\tensor{\overline{a}}{_C^B} = \delta^A_c$. The coefficients $\tensor{a}{^A_B}$ are functions of $x$, not $\Omega$.

\subsection*{a) Tensor transformations under group (\ref{Eq1}).} 

A vector $F$ with contravariant components $F^\rho$ ($\rho=1,2,3,4;$ $1,2,...,n$), and a vector $G$ with covariant components $G_\rho$ transforms as follows\footnote{[The usual partial derivatives are denoted by $\partial_i \equiv \partial/\partial x^i$, while the partial derivatives with respect to $\Omega$ are indicated by $\partial_{\Omega,A} \equiv \partial/\partial \Omega^A$. Translation note by A.H.]}:
\begin{align}\label{Eq2-3}
	F'^i &= F^i, \quad F'^A = \tensor{a}{^A_B}(x) F^B + (\partial_k \tensor{a}{^A_B}(x))\,\Omega^B\, F^k,\\
	G'^i &= G^i + (\partial_{i'} \tensor{\overline{a}}{_B^A}(x'))\, \Omega'^B\, G_A, \quad G'_A = \tensor{\overline{a}}{_A^B}(x') \,G_B.
\end{align} 
(It is enough to recall, for instance, that 
\begin{align*}
	F'^i = (\partial_k x'^i)\, F^k + (\partial_{\Omega,A} x'^i)\, F^A. \quad )
\end{align*}
One can show immediately that
\begin{align}\label{Eq4}
F^i G_i + F^A G_A = \text{invariant}.
\end{align}
(Recall the relation between the coefficients $a$ and $\overline{a}$:
\begin{align*}
(\partial_i \tensor{\overline{a}}{_B^A})\, \tensor{a}{^B_C} = -\tensor{\overline{a}}{_B^A}\,(\partial_i \tensor{a}{^B_C}). \quad )
\end{align*}

\subsection*{b) The metric tensor $g_{\rho\sigma}(x,\Omega)$.}

One introduces a symmetric metric tensor, whose components $g_{ik}$, $g_{iA}$, and $g_{AB}$ transforms in the following way:
\begin{equation}\label{Eq5}
\begin{split}
g'_{ik} &= g_{ik} + (\partial_i \,\tensor{\overline{a}}{_B^A}) \,\Omega'^B\, g_{Ak} +(\partial_k\,\tensor{\overline{a}}{_B^A})\, \Omega'^B\, g_{iA} 
+ (\partial_i\,\tensor{\overline{a}}{_C^A}) (\partial_k \,\tensor{\overline{a}}{_D^B})\, \Omega'^C\, \Omega'^D\, g_{AB},\\
g'_{iA} &= \tensor{\overline{a}}{_A^B} g_{iB} + (\partial_i\, \tensor{\overline{a}}{_C^B})\, \Omega'^C\, \tensor{\overline{a}}{_A^D}\, g_{BD}, \\
g'_{AB} &= \tensor{\overline{a}}{_A^C}\, \tensor{\overline{a}}{_B^D}\, g_{CD}.
\end{split}
\end{equation}	
Also, it may be convenient to introduce a condition of homogeneity, for instance of the type
\begin{equation}\label{Eq6}
g_{AB}\, \Omega^A\, \Omega^B =1.
\end{equation}
(In the ordinary space, it would correspond to spherical surfaces of unitary radius.)

\subsection*{c) Normal form of the line element.}

From general relativity, we know that, by taking an arbitrary point [of the spacetime], it is always possible to reduce its line element to the normal form. In our case, it corresponds to have
\begin{equation}\label{Eq7}
g_{iA} = 0; \quad g_{AB} = \tensor{\delta}{_A^B}; \quad g_{ik} = \mathring{g}_{ik}; \quad \mathring{g}_{ik} = \mathring{g}^{ik} = \text{diag}(1,1,1,-1).
\end{equation}
It is easy to convince oneself that this implies that
\begin{align}
	&g_{AB} \text{ is independent from  } \Omega;\tag{8a}\label{Eq8a}\\
	&g_{Ai} = f_{AB,i}(x)\Omega^B \text{, namely linearly dependent on  } \Omega.\tag{8b}\label{Eq8b}
\end{align}
Note that these properties are invariant with respect to the group of transformations (\ref{Eq1}). 

In the new reference frame, $g'_{AB} = \tensor{\delta}{_A^B}$. From the third relation in (\ref{Eq5}), it then follows that $\tensor{\overline{a}}{_A^C} \tensor{a}{^D_B} g_{CD} = \tensor{\delta}{_A^B}$. Moreover,
that $\overline{a}$ depends only on $x$ is possible only if $g_{CD}$ is independent of $\Omega$.

Thinking in a similar way, one finds that
\begin{align*}
\tensor{\overline{a}}{_A^B}\, g_{iB} + (\partial_i \tensor{\overline{a}}{_C^B})\, \Omega'^C\, \tensor{a}{^D_A}\, g_{BD} = 0.
\end{align*}
The second term is linear and homogeneous in $\Omega$, and the first as well hence.

\subsection*{d) The underlining operation of the indices.}

The following equations \textit{define} the new tensors with $A$ contravariant and $i$ covariant underlined indices (the reasons for introducing this notation will become clearer in the following):
\begin{equation}\tag{9}\label{Eq9}
\begin{split}
g^{il} g_{kl} &= \tensor{\delta}{^i_k}, \\
\tensor{g}{^{\underline{A}}^{\underline{B}}} \, g_{BC} &= \tensor{\delta}{^A_C},\\
F^{\underline{A}} &= F^A + \tensor{g}{^{\underline{A}}^{\underline{B}}}\, g_{Bk}\, F^k ,\\
G_{\underline{k}} &= G_k - \tensor{g}{^{\underline{A}}^{\underline{B}}}\, g_{Bk}\, G_A.
\end{split}
\end{equation}
Let us introduce the abbreviated notation:
\begin{equation}\tag{10}\label{Eq10}
	\tensor{g}{^{\underline{A}}_k} \equiv \tensor{g}{^{\underline{A}}^{\underline{B}}}\,\tensor{g}{_{\underline{B}}_k}.
\end{equation}
It is worth to remark that
\begin{equation}
	\tensor{g}{^{\underline{A}}_k} \neq \tensor{\delta}{^A_k}.
\end{equation}
Otherwise, $g_{AB}$ would be $\Omega-$dependent.

It is obvious how to transit from the underlining operation of vector indices to tensor ones: it is enough to think of it as a product of vectors.

One also easily gets that
\begin{align*}
\tensor{g}{_A_{\underline{k}}} = \tensor{g}{^{\underline{A}}^k} = 0.
\end{align*}

Beside $\tensor{g}{^{\underline{A}}^{\underline{B}}} g_{BC} = \tensor{\delta}{^A_C}$, the transformation law for $\tensor{g}{^{\underline{A}}^{\underline{B}}}$ results to be the reciprocal of the one for $\tensor{g}{_B_C}$:
\begin{align*}
(\tensor{g}{^{\underline{A}}^{\underline{B}}}){'} = \tensor{a}{^A_F} \,\tensor{a}{^B_G}\, \tensor{g}{^{\underline{F}}^{\underline{G}}}.
\end{align*}
It follows that
\begin{align*}
	(\tensor{g}{^{\underline{A}}_k})' = (\tensor{g}{^{\underline{A}}^{\underline{B}}}){'} (g_{Bk})' = \tensor{a}{^A_F} \,\tensor{g}{^{\underline{F}}^{\underline{G}}}\,g_{Gk} - (\partial_k \,\tensor{a}{^A_I})\,\Omega^I.
\end{align*}
It is easy now to conclude that
\begin{equation}\tag{11}\label{Eq11}
\begin{split}
F'^{\underline{A}} &= \tensor{a}{_B^A} \, F^{\underline{A}} \\
G'_{\underline{k}} &= \tensor{a}{_B^A} \, G_{\underline{k}} \\
F^{\underline{A}} G_A + F^k G_{\underline{k}} &= F^A G_A + F^k G_k.
\end{split}
\end{equation}
The lowering and lifting of an index can be, respectively, defined in the following way:
\begin{equation*}
\begin{split}
g_{BA}\, F^A + g_{Bk}\, F^k = F_B; \quad g_{iA} \, F^A + g_{ik}\,F^k = F_i.\\
g^{BA}\, F_A + g^{Bk}\, F_k = F^B; \quad g^{iA} \, F_A + g^{ik}\,F_k = F^i.
\end{split}
\end{equation*}
Together with (\ref{Eq11}), these relations make explicit the reason why the underlining operation was introduced.

Finally, it is not difficult to realize that the relations for $g$ with upper and lower indices, underlined or not, are such that, once ascribed the $ \tensor{g}{_{\underline{i}}_{\underline{k}}}$, $\tensor{g}{_i_A}$, and $\tensor{g}{_A_B}$, it is possible to recover the remaining ones. Note that, if at this point $\tensor{g}{_{\underline{i}}_{\underline{k}}} = \mathring{g}_{ik}$ (special relativity), the choice would be invariant under the group (\ref{Eq1}).

\subsection*{e) Covariant derivative in the $\omega-$space.}

The notation
\begin{equation}\tag{12}\label{Eq12}
D_{\underline{k}} \equiv \partial_{\underline{k}} - \tensor{g}{^{\underline{A}}_k} \, \partial_{\Omega,A}
\end{equation}
define an operation that is invariant under transformations of the $\omega-$space, and therefore the index $k$ is underlined. Remark that $G'_{\underline{k}} = G_{\underline{k}}$\,, and $G_{\underline{k}} = G_k - \tensor{g}{_k^{\underline{A}}} \,G_A$\,. It is enough, hence, to show that $\partial_k$ transforms as $G_k$, and $\partial_{\Omega,A}$ as $G_A$:
\begin{align*}
	\partial_{k'} &= \frac{\partial x^i}{\partial x'^{k}} \frac{\partial}{\partial x^i} 
	+ \frac{\partial \Omega^A}{\partial x'^{k}} \frac{\partial}{\partial \Omega^A}
	= \frac{\partial}{\partial x^k}  + \frac{\partial \tensor{\overline{a}}{_C^A}}{\partial x'^{k}} \,\Omega'^C\, \frac{\partial}{\partial \Omega^A}\,,\\
	\partial_{\Omega,A'} &= \frac{\partial\Omega^B}{\partial \Omega'^A}\frac{\partial}{\partial\Omega^B} + \frac{\partial x^i}{\partial \Omega'^A}\frac{\partial}{\partial x^i} = \tensor{\overline{a}}{_A^B}\frac{\partial}{\partial \Omega^B}\,.
\end{align*}

\subsection*{f) Field strength tensors.}

Acting with $D_{\underline{k}}$ over $\tensor{g}{^{\underline{A}}_k}$ (that is, in practice over $g_{Ai}$, which, among the components of the metric tensor, are the ones that establishes the connection between the $\omega-$space and the ordinary space), it is possible to construct an antisymmetric tensor in $i$ and $k$:
\begin{equation}\tag{13}\label{Eq13}
(\tensor{F}{^{\underline{A}}_{\underline{i}}_{\underline{k}}})' = D_{\underline{i}}\,\tensor{g}{^{\underline{A}}_k} - D_{\underline{k}}\,\tensor{g}{^{\underline{A}}_i}\, .
\end{equation}

The indices $A$, $i$, and $k$ are underlined, while the transformation law for the [field strength] tensor results to be the following:
\begin{equation}\tag{14}\label{Eq14}
(\tensor{F}{^{\underline{A}}_{\underline{i}}_{\underline{k}}})' = \tensor{a}{^A_B} \, \tensor{F}{^{\underline{B}}_{\underline{i}}_{\underline{k}}}
\end{equation}

Eq. (\ref{Eq14}) expresses that $F$ transforms as a vector with respect to the $\omega-$space index, and as an invariant with respect to the ordinary space indices. This is a direct consequence from the law of transformation satisfied by $\tensor{g}{^{\underline{A}}_k}$:
\begin{equation}\tag{15}\label{Eq15}
(\tensor{g}{^{\underline{A}}_k})' = \tensor{a}{^A_B} \,\tensor{g}{^{\underline{B}}_k} - (\partial_k\, \tensor{a}{^A_C})\, \Omega^C.
\end{equation}

The form and behavior of $\tensor{F}{^{\underline{A}}_{\underline{i}}_{\underline{k}}}$ with respect to the spacial axis justify its nomenclature as a \textit{[field strength] tensor}\footnote{[In the original, \enquote{tensore delle forze}. Translation note by A.H.]}, and the consideration of Eq. (\ref{Eq15}) as a sort of \textit{gauge transformation} induced on the \enquote{potentials} $\tensor{g}{^{\underline{A}}_k}$ (or $g_{Ak}$) by the transformations (\ref{Eq1}).

Notice that if the $\omega-$space is assumed to be 2-dimensional and homogeneous, with $g_{AB} = \tensor{\delta}{_A^B}$, then Eq. (\ref{Eq6}) becomes
\begin{align*}
(\Omega_1)^2 + (\Omega_2)^2 = 1. 
\end{align*}
Setting $\Omega_1 = \cos(x_5)$, and $\Omega_2 = \sin(x_5)$, it is easy to see that the group of Klein and Kaluza,
\begin{align*}
x'_5 = x_5 + f(x_1,...,x_4)
\end{align*}
is contained in (\ref{Eq1}). Indeed,
\begin{align*}
\cos(x'_5) = \cos(x_5 + f) = \cos (f(x))\,\cos(x_5) - \sin(f(x)) \,\sin(x_5),
\end{align*}
[and then]
\begin{align*}
	\Omega'_1 = a(x)\,\Omega_1 + b(x)\,\Omega_2.
\end{align*}

Furthermore, we want to stress explicitly the close analogy between the transformation laws of $g_{Ai}$ and the gauge transformations of the electromagnetic potentials, writing it for the 1-dimensional case,
\begin{align*}
g'_{Ai} = \overline{a}_A\,g_{Ai} + (\partial_i\,\overline{a}_A)\,\Omega'^A\,\overline{a}_A\,g_{AA} \quad (\text{no sum}).
\end{align*}
It is easy to realize that, being $g_{Ai}$ linear and homogeneous in $\Omega$, we can write
\begin{equation}\tag{16}\label{Eq16}
\tensor{F}{^{\underline{A}}_{\underline{i}}_{\underline{k}}} = \tensor{f}{^{\underline{A}}_{B,}_{\underline{i}}_{\underline{k}}}\,\Omega^B.
\end{equation}
With a simple calculation, one finds for the coefficients $f$ the following expression:
\begin{equation}\tag{17}\label{Eq17}
\tensor{f}{^{\underline{A}}_{B,}_{\underline{i}}_{\underline{k}}} = \partial_i\,\tensor{f}{^{\underline{A}}_{B,}_{k}} - \partial_k\, \tensor{f}{^{\underline{A}}_{B,}_{i}} - \tensor{f}{^{\underline{A}}_{C,}_{k}}\,\tensor{f}{^{\underline{C}}_{B,}_{i}} 
+ \tensor{f}{^{\underline{A}}_{C,}_{i}}\,\tensor{f}{^{\underline{C}}_{B,}_{k}},
\end{equation}
with $\tensor{f}{^{\underline{A}}_{B,}_{k}}$ (see Eqs (\ref{Eq8b}) and (\ref{Eq10})) defined as
\begin{align*}
\tensor{g}{^{\underline{A}}_{k}} = \tensor{f}{^{\underline{A}}_{B,}_{k}}\,\Omega^B, \quad \tensor{f}{^{\underline{A}}_{B,}_{k}} = \tensor{g}{^{\underline{A}}^{\underline{C}}}\,\tensor{f}{_{C}_{B,}_{k}}.
\end{align*}
Beside tensor (\ref{Eq17}), it is possible to define another one equivalent to it, with all indices lowered. One shall note the formal resemblance between $\tensor{f}{^{\underline{A}}_{B,}_{\underline{i}}_{\underline{k}}}$ and the Riemann tensor. Well, it is possible to show that this [feature] attribute to the $\omega-$space a role analogue to the one played by the Riemann tensor to the chronotope. In fact, its vanishing is a necessary and sufficient condition for $g_{Ai}=0$ at \textit{every} point. It is obvious that the condition is a necessary one. And it is intuitive, according to the interpretation adopted, that if the [field strength] vanishes, its potentials result constant. We omitted here the proof.

At this point, it is opportune to anticipate that, in order to have $g_{AB} = \tensor{\delta}{_A^B}$ \textit{everywhere}, another condition is required (see further).

\subsection*{g) The scalar potential.}\label{sec.II-g}

Let us consider the expression
\begin{align*}
\tensor{\Psi}{_{\underline{i}}_{,(A}_{B)}} \equiv \frac{1}{2} (f_{i,AB} + f_{i,BA} - \partial_i\,g_{AB}), 
\end{align*}
constructed with the coefficients of the potentials (see Eq. (\ref{Eq8b})). This [expression] is symmetric in $A,B$, and th index $\underline{i}$ is underlined, one $\Psi$ transforms with respect to (\ref{Eq1}) as a tensor, as it can easily be shown.

Let us define the antisymmetric quantity
\begin{align*}
\tensor{\Psi}{_{\underline{i}}_{,[A}_{B]}} \equiv \frac{1}{2} (f_{i,AB} - f_{i,BA}).
\end{align*}
It holds then
\begin{equation}\tag{18}\label{Eq18}
f_{i,AB} = \frac{1}{2} \partial_i\,g_{AB} +\tensor{\Psi}{_{\underline{i}}_{,(A}_{B)}}+ \tensor{\Psi}{_{\underline{i}}_{,[A}_{B]}}.
\end{equation}

We decomposed the coefficients $f_{i,AB} $ in its symmetric and antisymmetric parts: one easily recognize that such decomposition is invariant under the group (\ref{Eq1}).

Now, there are two possibilities:
\begin{itemize}
	\item[1)] if one takes $\tensor{\Psi}{_{\underline{i}}_{,(A}_{B)}} = 0$ (this hypothesis is invariant), the [strength fields] result antisymmetric not only in $i,k$, but also in $A,B$.
	
	\item[2)] setting, in general, $\Psi$ as the covariant derivative (see further) of a symmetric function in $A,B$,
	\begin{align*}
	\tensor{\Psi}{_{\underline{i}}_{,(A}_{B)}} = \tensor{\Phi}{_{(A}_{B);}_{\underline{i}}}\;\;,
	\end{align*}
	one would introduce a scalar field $\Phi_{AB}$, under Lorentz group, with isotopic spin equals to two, beside $f_{i,AB}$, which transforms as a vector.
\end{itemize}

Hypothesis 1) is not necessary, but it has the merit of simplifying considerably the metric. In fact, if 1) holds and the tensor (\ref{Eq17}) vanishes, then Eq. (\ref{Eq18}) reduces to $\partial_i\,g_{AB}=0$, that is, it is possible to set $g_{AB} =\tensor{\delta}{_A^B}$ not only at one [arbitrary] point, but for the entire space.

\section{Dirac's equation in the chronotope}

In special relativity, the matrices $\mathring{\gamma}_\mu$ may be deduced from the metric tensor by means of the relations
\begin{equation}\tag{I}\label{EqI}
\begin{split}
\mathring{\gamma}_\mu\,\mathring{\gamma}_\nu + \mathring{\gamma}_\nu\,\mathring{\gamma}_\mu &= 2 \mathring{g}_{\mu\nu}\cdot I\,,\quad (\mu,\nu=1,2,3) \\
(\mathring{\gamma}_4)^2 &= \mathring{g}_{44} = -1.
\end{split}
\end{equation}
The last relation suggests that $\mathring{\gamma}_4$ results antihermitian. The generalization of (\ref{EqI}) to general relativity is carried on by posing
\begin{equation}\tag{II}\label{EqII}
\gamma_i\,\gamma_k + \gamma_k\,\gamma_i = 2g_{ik}(x)\,, \quad (i,k=1,2,3,4).
\end{equation}
The solutions $\gamma_i$ of (\ref{EqII}) are linear combinations of $\mathring{\gamma}_i$ with point-dependent coefficients.

By means of a non singular matrix $S$, one may go from one solution of (\ref{EqII}) to another solution 
\begin{align*}
\gamma'_i = S^{-1}\,\gamma_i\,S.
\end{align*}
Taking only infinitesimal transformations $S= I + \varepsilon\,\Lambda$, and writing $\gamma'$ in the form $\gamma' = \gamma + \varepsilon\,\eta$\,, one finds that it must hold
\begin{align*}
\eta_i = \gamma_i\,\Lambda - \Lambda\,\gamma_i\,.
\end{align*} 
By using this relation and recalling that, besides (\ref{EqII}),
\begin{align*}
g_{ik;l} \equiv \partial_l\,g_{ik} - \left\{\mqty{&r&\\i& &l}\right\}\,g_{rk} - \left\{\mqty{&r&\\k& &l}\right\}\,g_{ir} = 0, 
\end{align*}
one obtains that there exists four operators $\Lambda_l$, and that it is possible to define the covariant derivative of $\gamma$ as in the following:
\begin{equation}\tag{III}\label{EqIII}
\gamma_{i;l} \equiv \partial_l\,\gamma_{i} -  \left\{\mqty{&r&\\i& &l}\right\}\,\gamma_r + \Lambda_l\,\gamma_i - \gamma_i\,\Lambda_l = 0.
\end{equation}
These equations might be useful to define $\Lambda$, unless diagonal matrices,
\begin{align*}
\Lambda' = \Lambda + c\,I.
\end{align*}

From the tensor derivative formulas, it is now possible to recover the respective ones for the spinors:
\begin{align*}
\tensor{s}{^k_{;l}} \equiv \partial_l\,s^k +\left\{\mqty{&k&\\r& &l}\right\}\,s^r = (\overline{\Psi}\,\gamma^k\,\Psi )_{;l} = \overline{\Psi}_{;l}\,\gamma^k \,\Psi + \overline{\Psi}\,\gamma^k\,\Psi_{;l}\;.
\end{align*}

For the last step, it has been taken into account Eq. (\ref{EqIII}). By comparing the first and the last expressions, one finds easily that
\begin{equation}\tag{IV}\label{EqIV}
\Psi_{;l} \equiv \partial_l\,\Psi + \Lambda_l\,\Psi\;,\quad \overline{\Psi}_{;l} = \partial_l\,\overline{\Psi} -\overline{\Psi}\,\Lambda_l\;.
\end{equation}
(The second is a consequence of the first.)

In particular, it follows that
\begin{align*}
(\overline{\Psi}\,\Psi)_{;l} = \partial_l\,(\overline{\Psi}\,\Psi)\;.
\end{align*}

The most natural generalization of Dirac's equation is the following:
\begin{equation}\tag{V}\label{EqV}
\begin{split}
\gamma^k\,\qty( \Psi_{;k} - \frac{ie}{\hbar}\,\phi_k\,\Psi ) +m\Psi = 0\;,\\
\qty(\overline{\Psi}_{;k} + \frac{ie}{\hbar}\,\phi_k \,\overline{\Psi})\gamma^k - m\Psi =0\;.
\end{split}
\end{equation}

One may see that these equations can always be \textit{locally} reduced to the ordinary Dirac's equations; in the iterated equations, instead, the Riemann tensor appears explicitly.

The continuity equation that follows from (\ref{EqV}) may be written as
\begin{align*}
(\overline{\Psi}\,\gamma^k\,\Psi )_{;k} = \partial_k\,(\sqrt{g}\,\overline{\Psi}\,\gamma^k\,\Psi)=0.
\end{align*}

\section{Dirac's equation in the $\omega-$space}

For the extension of the previous results to the $\omega-$space, we proceed in the following way.

To the Eqs. (\ref{EqII}), we set the correspondence:
\begin{align*}
\gamma_i\,\gamma_k + \gamma_k\,\gamma_i &= 2g_{ik}\,, \\
\gamma_i\,\gamma_A + \gamma_A\,\gamma_i &= 2g_{iA}\,, \\
\gamma_A\,\gamma_B + \gamma_B\,\gamma_A &= 2g_{AB}\,.
\end{align*}

Defining
\begin{align*}
\gamma_{\underline{k}} = \gamma_k - \tensor{g}{^{\underline{A}}_k}\,\gamma_A\,,\quad \gamma^i = g^{ik}\,\gamma_{\underline{k}}\,,\quad \gamma^{\underline{A}} = \tensor{g}{^{\underline{A}}^{\underline{B}}}\,\gamma_B\,,
\end{align*}
and recalling that $\tensor{g}{^{\underline{A}}^i}=0$, one obtain the equivalent equations
\begin{equation}\tag{II'}\label{EqII'}
\begin{split}
\gamma^i\,\gamma^k + \gamma^k\,\gamma^i &= 2g^{ik}\,, \\
\gamma^{\underline{A}}\,\gamma^{\underline{B}} + \gamma^{\underline{B}}\,\gamma^{\underline{A}} &= 2\tensor{g}{^{\underline{A}}^{\underline{B}}}\,,\\
\gamma^{\underline{A}} \,\gamma^i + \gamma^i\,\gamma^{\underline{A}} &= 0.
\end{split}
\end{equation}
Let us denote by $x^\rho$ the coordinates at the point $(x^i,\Omega^A)$, and use the convention of upper and lower underlined indices. One finds, then, the following equations:
\begin{align}
\tensor{\gamma}{^\mu_{;\rho}} &\equiv \partial_\rho\,\gamma^\mu + \left\{\mqty{&\mu&\\\rho& &\sigma}\right\}\,\gamma^\sigma + \Lambda_\rho\,\gamma^\mu - \gamma^\mu\,\Lambda_\rho = 0\,, \tag{III'}\label{EqIII'} \\
\Psi_{;\rho} &\equiv D_\rho\,\Psi + \Lambda_\rho\,\Psi\,,\quad \overline{\Psi}_{;\rho} = D_\rho\,\overline{\Psi} - \overline{\Psi}\,\Lambda_\rho\,,
\tag{IV'}\label{EqIV'}
\end{align}
where
\begin{align*}
D_{\underline{i}} = \partial_{\underline{i}} - \tensor{g}{^{\underline{A}}_i}\, D_A\,,\quad D_A \equiv \partial_{\Omega,A}\,.
\end{align*}

One arrives at Eqs. (\ref{EqIII'}) by remembering that $\tensor{g}{^{\underline{A}}^{\underline{B}}}$ do not depend on $\Omega$, and by observing that, with say,
\begin{align*}
\left\{\mqty{&\underline{A}&\\B& &i}\right\} = \frac{1}{2}\tensor{g}{^{\underline{A}}^{\underline{C}}}(\partial_i\,g_{BC} + D_B\,g_{Ci} - D_C\,g_{Bi})\,,
\end{align*}
expressions like
\begin{align*}
\tensor{\Gamma}{_{\underline{i}}^{\underline{A}}} \equiv \partial_i\,\gamma^{\underline{A}} + \left\{\mqty{&\underline{A}&\\B& &i}\right\}\,\gamma^{\underline{B}} 
\end{align*}
transforms as vectors in the $\omega-$space.

A more detailed investigation would show that Eqs. (\ref{EqIII'}), which define the operators $\Lambda_\rho$\,, can be reduced to the following equations:
\begin{align*}
\partial_i\,\gamma^{\underline{A}} + \left\{\mqty{&\underline{A}&\\B& &i}\right\}\,\gamma^{\underline{B}} + \Lambda_{\underline{i}} \,\gamma^{\underline{A}} - \gamma^{\underline{A}}\,\Lambda_{\underline{i}} = 0\,,\\
\Lambda_{\underline{i}}\,\gamma^k - \gamma^k\,\Lambda_i = 0\,,\quad \Lambda_A = 0.
\end{align*}

In particular, as you can see, it results possible to set $\Lambda_A$ equals to zero. This helps to simplify significantly the calculation. Dirac's equation results, in this way, in the form 
\begin{equation}\tag{V'}\label{EqV'}
\qty[\gamma^k\,(D_{\underline{k}} + \Lambda_{\underline{k}})\, + \gamma^{\underline{A}}\, D_A\, + M]\,\Psi = 0.
\end{equation}

Finally, we want to make some observations on the $\gamma$ matrices. It results that the $\gamma^{\underline{A}}$ are only dependent on the $x^i-$coordinates, and not on the $\Omega^A$, being $\tensor{g}{^{\underline{A}}^{\underline{B}}}$ functions only of $x^i$. Once $\gamma^{\underline{A}}$ anticommute with all $\gamma^i$, [it holds that] they are proportional to the $\gamma^5$. If we restrict [the analysis] to the 3-dimensional, homogeneous $\omega-$space, we can write 
\begin{align*}
\gamma^{\underline{A}} = \tensor{\alpha}{^A_B}(x)\,\tau^B\,\gamma^5\,,
\end{align*} 
where $\tau^B$ are the ordinary matrices of isotopic spin,
\begin{align*}
\mathring{\tau}^A\,\mathring{\tau}^B + \mathring{\tau}^B\,\mathring{\tau}^A = 2\tensor{\delta}{_A^B}\,,\quad [\mathring{\tau}^A,\gamma^i] = 0.
\end{align*}
In this case, $\gamma^{\underline{A}}$ are $8\times8$ matrices.

\section{Lagrangian formulation and conservation theorems}

Dirac's equation can be deduced from the variation of the following Lagrangian function:
\begin{align*}
L_D &= \int dx\,d\Omega\,\sqrt{g}\, \mathcal{L}_D\\
&= \int dx\,d\Omega\,\sqrt{g}\, \overline{\Psi} \qty[\gamma^k\,(\partial_k - \tensor{g}{^{\underline{A}}_{k}}\,D_A + \Lambda_{\underline{k}} ) + \gamma^{\underline{A}}\,D_A + M ]\Psi\,+ \,\text{h.c.}
\end{align*}
The interaction potentials are contained in $\tensor{g}{^{\underline{A}}_{k}}$.

The equations of motion for these potentials may be deduced from a total Lagrangian obtained by adding to $L_D$ the invariant terms constructed with the same potentials.

Let us consider the [field strength] tensors $\tensor{f}{^{\underline{A}}_{B,}_{\underline{i}}_{\underline{k}}}$\,, from which it is possible to construct the invariant
\begin{align*}
\tensor{f}{^{\underline{A}}_{\underline{B},}_{\underline{i}}_{\underline{k}}}\cdot \tensor{f}{^i^k_A_B}\,.
\end{align*}

However, it results impossible to introduce a mass term of the type
\begin{align*}
\mu\,g_{\underline{i}A}\,g^{i\underline{A}}\,,
\end{align*}
as the (necessary) underlining of the indices makes inevitable the presence of $g_{\underline{i}A}$ and $g^{i\underline{A}}$, which, as we know, are identically zero.

We shall, therefore, write
\begin{align*}
	L = \int dx\,d\Omega\,\sqrt{g}\, \qty( \mathcal{L}_D - \frac{1}{4}\,\tensor{f}{^{\underline{A}}_{\underline{B},}_{\underline{i}}_{\underline{k}}}\cdot \tensor{f}{^i^k_A_B}\,)\,.
\end{align*}
From the variation of $L$, one may deduce not only the equations of motion, as the continuity equations related to the invariance of $L$ under the group of transformations (\ref{Eq1}).

In order to simplify the calculations, one may set, for the [particular] case when $g_{AB}=\tensor{\delta}{_A^B}$ everywhere (see the paragraph g) in section \ref{sec.II-g}),
\begin{align}
	f_{i,AB} = \phi_{i,[AB]}\,.
\end{align}
The strength field tensor results, in this case, antisymmetric in $i,k$ and in $A,B$ as well. Also, $\sqrt{g}=1$.

The equations of motion for $\phi_{i,AB}$ are the following:
\begin{equation}\tag{+}\label{Eq+}
\begin{split}
\frac{\partial^2 \phi_{i,AB}}{\partial x^{m^2}} - \frac{\partial^2 \Psi_{m,AB}}{\partial x^{m}\,\partial x^i} =\;& \frac{\partial}{\partial x^m} (\phi_{i,AC} \,\phi_{m,CB} - \phi_{m,AC}\,\phi_{i,CB}) \\
&+ \tensor{f}{^i^l_C_A}\,\phi_{l,CB} - \tensor{f}{^i^l_C_B}\,\phi_{l,CA} \\
&+ \frac{1}{2} \overline{\Psi}\,\gamma^i \qty(\Omega^B\, D_A - \Omega^A\,D_B )\Psi\,.
\end{split}
\end{equation}

The continuity equations holds
\begin{equation}\tag{++}\label{Eq++}
\begin{split}
 \frac{\partial}{\partial x^k}\Big\{\overline{\Psi}\,\gamma^k &\Big[ \qty(\Omega^A\, D_B - \Omega^B\,D_A ) 
 + \frac{1}{2} \gamma^{\underline{A}}\,\gamma^{\underline{B}} \Big]\Psi\, \\
 &+ 2(\phi_{i,AC}\,\tensor{f}{^i^k_B_C} - \phi_{i,BC}\,\tensor{f}{^i^k_A_C} )\Big\} + D_A\,Q^{\underline{A}} = 0.
\end{split}
\end{equation}
The expression for $Q^{\underline{A}}$, which we shall dismiss, is more involved; for it is a total divergence after integration over the $\Omega^A-$variables.

Taking the derivative of Eq. (\ref{Eq+}) with respect to $x^i$, and performing the sum over [the index] $i$, one obtains the following identity:
\begin{align*}
\partial_i \, \qty[ \overline{\Psi}\,\gamma^i \qty(\Omega^B\, D_A - \Omega^A\,D_B )\Psi 
+ 2(\tensor{f}{^i^l_C_A}\,\phi_{l,CB} - \tensor{f}{^i^l_C_B}\,\phi_{l,CA} ) ] = 0,
\end{align*}
which, compared with Eq. (\ref{Eq++}) after integration over $\Omega$, allow us to conclude that
\begin{align*}
\partial_i \int d\Omega\,\overline{\Psi}\,\gamma^i\frac{\gamma^{\underline{A}}\,\gamma^{\underline{B}}}{2}\, \Psi = 0.
\end{align*}
To these continuity equations, valid in the ordinary space, corresponds similarly the conservation theorems.

\section{Concluding remarks}

The theory presented in the previous sections fulfills, as it can be seen, its purpose to deduce the form of the baryon coupling with the meson field from the hypothesis of invariance with respect to a certain group of transformations, thereby avoiding the need to postulating it directly, as done by Pais.

The meson field provided by this theory, however, is a vector field in the ordinary space. The considerations developed in [paragraph] g) of section \ref{sec.II-g} show that one may also introduce a scalar field (and a pseudoscalar as well), but at the price of an [additional] hypothesis that appear somewhat artificial, and do not exclude the concomitant presence of the vector field. Besides, the mass of the meson, once it cannot be introduced in the Lagrangian, should follows from the interaction of the meson field with other fields.

Finally, we would like to remark that the electromagnetic field cannot be deduced from this theory in the same way as the meson field, but must be introduced as in the ordinary formulations.




\end{document}